\documentclass[twoside,english,linesnumbered, ruled, vlined, boxed]{sig-alternate}
\usepackage[T1]{fontenc}
\usepackage[latin9]{inputenc}
\usepackage{babel}
\usepackage{float}
\usepackage{calc}
\usepackage{algorithm2e}
\usepackage{amsmath}
\usepackage{graphicx}
\usepackage[unicode=true,pdfusetitle,
 bookmarks=true,bookmarksnumbered=false,bookmarksopen=false,
 breaklinks=false,pdfborder={0 0 1},backref=false,colorlinks=false]
 {hyperref}

\makeatletter

\providecommand{\tabularnewline}{\\}

\usepackage{fixref} 

\usepackage{enumitem} 
\setlist{nolistsep}

\usepackage{balance} 

\numberwithin{equation}{section}
\numberwithin{figure}{section}

\usepackage{tikz}
\usepackage{pgfplots}
\def\@copyrightspace{\relax}

\makeatother

\begin{document}

\title{Survey of resampling techniques for improving classification performance
in unbalanced datasets}

\author{Ajinkya More (ajinkya@umich.edu)}
\maketitle
\begin{abstract}
\emph{A number of classification problems need to deal with data imbalance
between classes. Often it is desired to have a high recall on the
minority class while maintaining a high precision on the majority
class. In this paper, we review a number of resampling techniques
proposed in literature to handle unbalanced datasets and study their
effect on classification performance.}
\end{abstract}

\section{Introduction}

Classification problems often suffer from data imbalance across classes.
This is the case when the size of examples from one class is significantly
higher or lower relative to the other classes. For many such problems
it is desirable to build classifiers with good performance on the
minority class. Using out of the box classifiers for such problems
may lead to suboptimal results with respect to this objective. 

In this paper we study several techniques for boosting classification
performance in the presence of data imbalance. We begin with examples
of some domains where unbalanced datasets is the norm.

\subsection{Examples}

\subsubsection{Fraud detection}

Detecting fraud in online transactions is a problem of significant
monetary impact. The number of fraudulent transactions is typically
a small fraction of all transactions and hence this problem is often
cited as a protypical data imbalance problem. In many cases, a fraud
detection system will flag potentially fraudulent transactions to
be reviewed manually by an analyst. Given the financial implication
of green lighting a fraudulent transaction, it is desirable to have
a classifier that can achieve near perfect recall on the fraudulent
class at the expense of lower precision, especially in the case when
the cost of manual review is much smaller.

\subsubsection{Product categorization}

E-commerce retailers categorize their product catalog into functional
groups to aid search retrieval. There is substaintial variation in
the number of items belonging to each category. For instance, there
are only a few iPhone models while the number of iPhone accessories
(e.g. cases, chargers, stylii, etc) is several hundred fold more.
There is bound to be a significant amount of overlap in the description
and images of items from these two categories. An automatic product
categorization system can potentially confuse between the two classes.
If the retailer is optimizing for revenue, it will be better to ensure
all iPhones are categorized correctly at the risk of classifying a
few iPhone accessories as iPhones. 

\subsubsection{Disease diagonsis}

For any given disease, the fraction of healthy people outnumber those
affected with it. In case of rare diseases, it is a tautology to say
that the dataset is highly imbalanced. If an automated classification
system is used to predict the presence of the disease (likely followed
by an expert evaluation), it is extremely important to have recall
on the disease class to be as close to 1 as possible. In this particular
case, even high precision on the minority class is essential since
a significant amount of expert analysis may be needed for avoiding
false positive disease prediction on healthy people.

\section{Notation and metrics}

Let us fix some notation to use in the remainder of the paper. We
will compare several methods for handling unbalanced datsets via a
case study on a synthetic two class dataset. Let us denote the majority
class by $L$ and the minority class by $S$. By these symbols, we
will refer to both the sets representing these classes as well as
the respective class labels. Denote by $r=|S|/|L|$ the ratio of the
size of the minority class to the majority class. Let the training
set be denote by $T$.

We will compare various techniques with respect to their effect on
the recall on the minority class $S$ and the precision on the majority
class $L$. This is motivated by applications to problems with the
following characteristics:
\begin{enumerate}
\item A large number of instances need to be evaluated.
\item The minority class is present in a small fraction of the instances.
\item Only instances flagged as minority class (by an automated classification
system) will be reviewed manually.
\item The cost of manual review is significantly lower relative to the cost
of a missed detection of the minority class. 
\end{enumerate}
This may in case in problems such as fraud detection, identifying
imminent hardware or software failures in large computer networks,
identifying product issues from online reviews, etc.

\section{Dataset}

We compare the performance of the various methods on a synthetic dataset.
We generate the dataset using the make\_classification function from
the Python library scikit-learn. We use the following parameters: 
\begin{enumerate}
\item n\_samples $=10000$ (number of data points)
\item n\_classes $=2$ (number of classes)
\item weights $=[0.1,0.9]$ (fraction of sizes of each class)
\item class\_sep $=1.2$ (the amount separation between the clusters defining
the two classes)
\item n\_features $=5$ (number of features)
\item n\_informative $=3$ (number of informative features)
\item n\_redundant $=1$ (number of redundant features)
\item n\_clusters\_per\_class $=1$
\end{enumerate}
For ease of visualization, we perform dimensionality reduction via
principal component analysis and pick the first two principal components
to form our dataset. A scatterplot for the original dataset is shown
below.
\begin{center}
\fbox{\begin{minipage}[t]{0.97\columnwidth}%
\includegraphics[width=0.99\columnwidth]{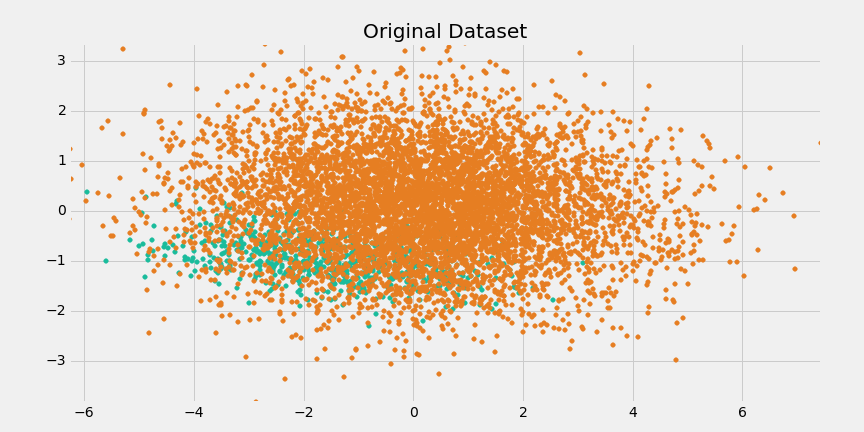}%
\end{minipage}}
\par\end{center}

\section{Method Comparison}

In this section we explore several methods for handling the data imbalance.
We split our dataset as 70\% training and 30\% test. We perform 5-fold
cross validation on the training set to select the best parameters
and report the results on the test set. The results were obtained
using python libraries scikit-learn and imbalanced-learn.

\subsection{Baseline}

We obtain baseline results using logistic regression where we perform
5-fold cross validation to search for the best regularization parameter
and the penalty type ($l_{1}$ or $l_{2}$). The classification boundary
on the training set is shown below.
\begin{center}
\fbox{\begin{minipage}[t]{0.97\columnwidth}%
\includegraphics[width=0.99\columnwidth]{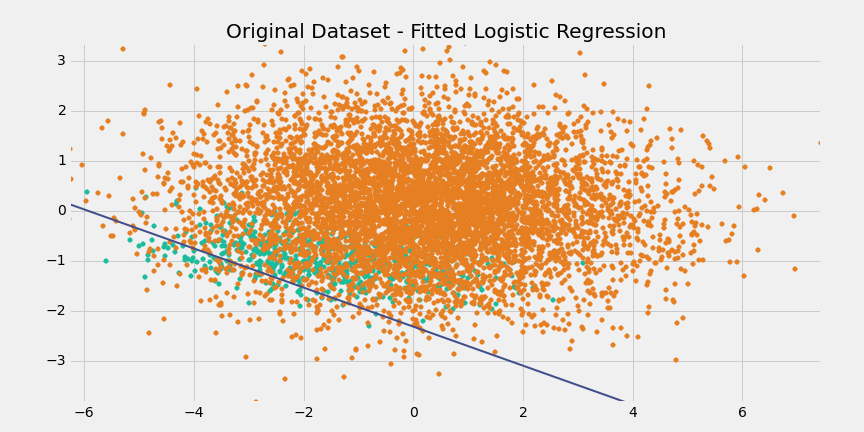}%
\end{minipage}}
\par\end{center}

The performance on the test set is as follows.
\begin{center}
\begin{tabular}{|c|c|}
\hline 
precision on $L$ & recall on $S$\tabularnewline
\hline 
\hline 
0.90 & 0.12\tabularnewline
\hline 
\end{tabular}
\par\end{center}

\subsection{Weighted loss function}

One technique for handling class imbalance, is to use a weighted loss
function. In order to boost performance on the minority class, the
penalty for misclassifying minority class examples can be increased.
For example the loss function for logistic regression is 
\[
-\Sigma_{j\in\mathcal{C}}\Sigma_{y_{i}=j}ln(P(y_{i}=j|x_{i};\theta)
\]
where $\mathcal{C}$ is the set of classes, $(x_{i},y_{i})$ is an
input-label pair in the training set and $\theta$ is the set of parameters.
A weighted loss function may be obtained as \cite{king2001logistic}
\[
-\Sigma_{j\in\mathcal{C}}\Sigma_{y_{i}=j}w_{j}ln(P(y_{i}=j|x_{i};\theta)
\]

In scikit-learn, this can be done for supported classifiers using
the '\emph{class\_weight}' parameter. Setting this parameter to '\emph{balanced'},
weights inversely proportional to the class sizes are used to multiply
the loss function.

The resulting decision boundary and the performance on the test set
are shown below.
\begin{center}
\fbox{\begin{minipage}[t]{0.97\columnwidth}%
\includegraphics[width=0.99\columnwidth]{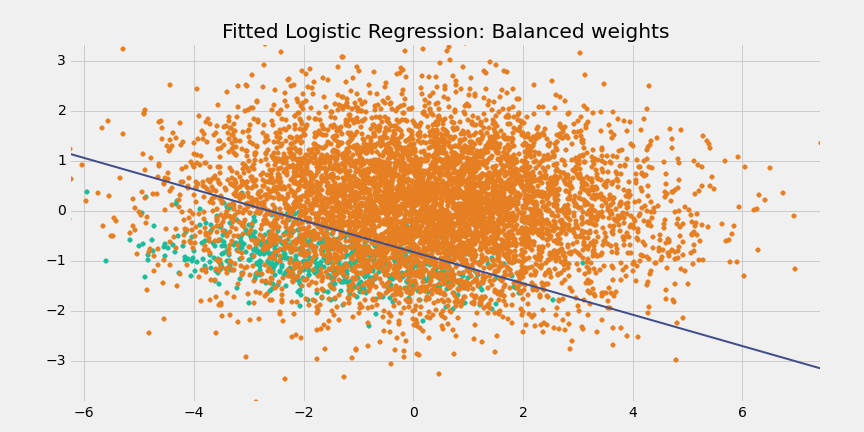}%
\end{minipage}}
\par\end{center}

\begin{center}
\begin{tabular}{|c|c|}
\hline 
precision on $L$ & recall on $S$\tabularnewline
\hline 
\hline 
0.98 & 0.89\tabularnewline
\hline 
\end{tabular}
\par\end{center}

\subsection{Undersampling methods}

\subsubsection{Random undersampling of majority class}

A simple undersampling technique is uniformly random undersampling
of the majority class. This can potentially lead to loss of information
about the majority class. However, in cases where each example of
the majority class is near other examples of the same class, this
method might yield good results. 
\begin{center}
\fbox{\begin{minipage}[t]{0.97\columnwidth}%
\includegraphics[width=0.99\columnwidth]{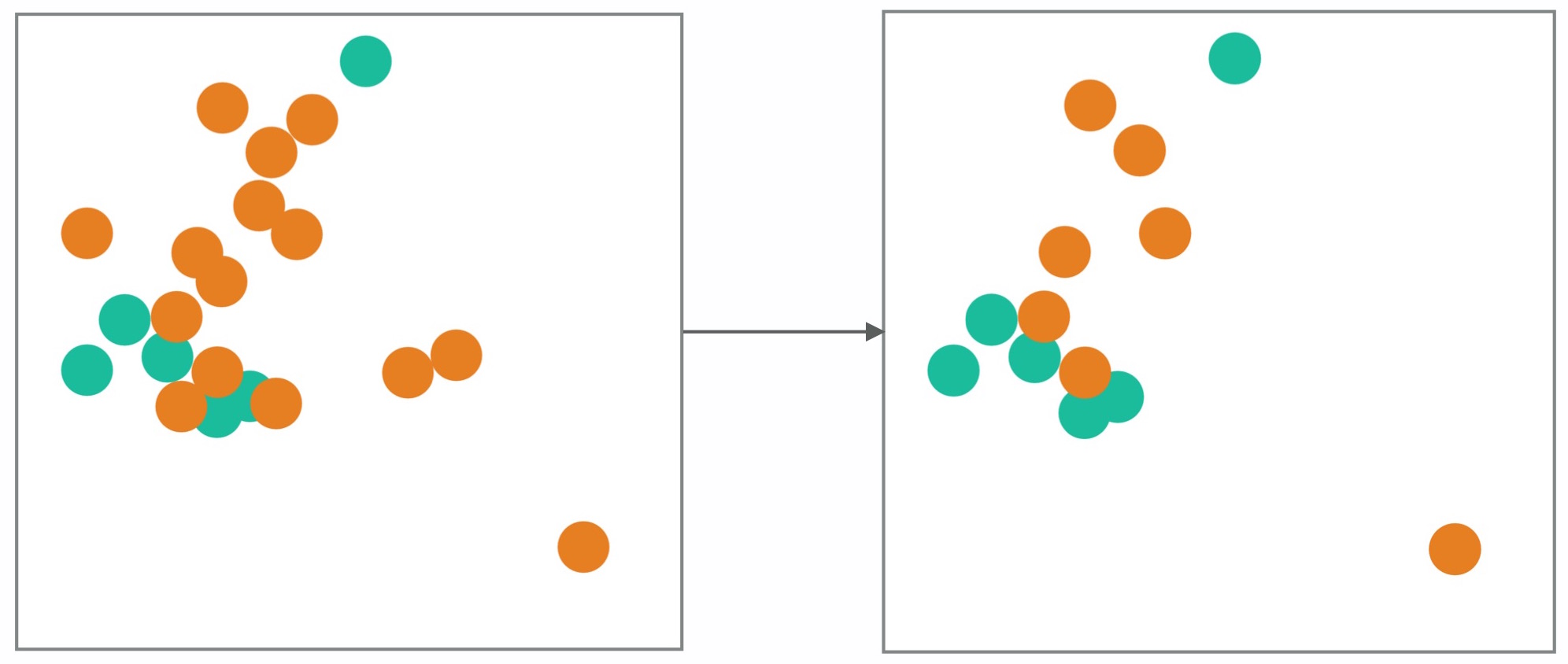}%
\end{minipage}}
\par\end{center}

We perform random undersampling of $L$ in order to achieve a value
of $r=0.5$. Fitting a logistic regression classifier to this resampled
dataset, we get the following performance.
\begin{center}
\begin{tabular}{|c|c|c|}
\hline 
 & $|L|$ & $|S|$\tabularnewline
\hline 
\hline 
Before resampling & 6320 & 680\tabularnewline
\hline 
After resampling & 1360 & 680\tabularnewline
\hline 
\end{tabular}
\par\end{center}

\begin{center}
\fbox{\begin{minipage}[t]{0.97\columnwidth}%
\includegraphics[width=0.99\columnwidth]{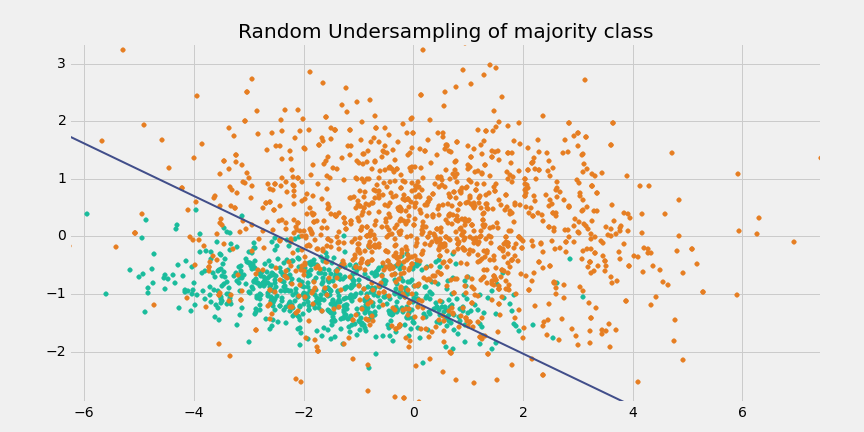}%
\end{minipage}}
\par\end{center}

\begin{center}
\begin{tabular}{|c|c|}
\hline 
precision on $L$ & recall on $S$\tabularnewline
\hline 
\hline 
0.97 & 0.82\tabularnewline
\hline 
\end{tabular}
\par\end{center}

\subsubsection{NearMiss-1}

The NearMiss family of methods \cite{nm} perform undersampling of
points in the majority class based on their distance to other points
in the same class. We discuss the 3 variants proposed in the paper
here.

In NearMiss-1, those points from $L$ are retained whose mean distance
to the $k$ nearest points in $S$ is lowest, where $k$ is a tunable
hyperparameter.
\begin{center}
\fbox{\begin{minipage}[t]{0.97\columnwidth}%
\includegraphics[width=0.99\columnwidth]{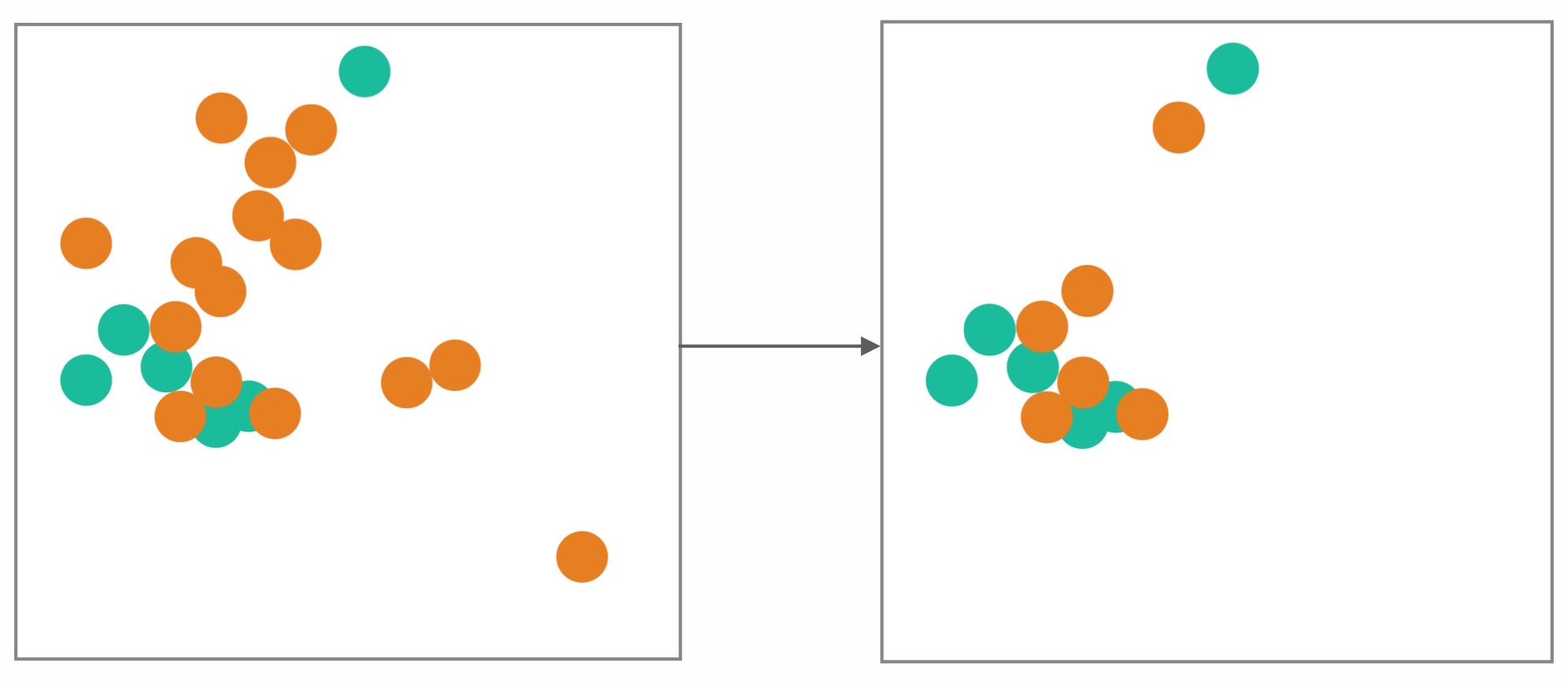}%
\end{minipage}}
\par\end{center}

We show below the result of using NearMiss-1 with $k=3$.
\begin{center}
\begin{tabular}{|c|c|c|}
\hline 
 & $|L|$ & $|S|$\tabularnewline
\hline 
\hline 
Before resampling & 6320 & 680\tabularnewline
\hline 
After resampling & 1360 & 680\tabularnewline
\hline 
\end{tabular}
\par\end{center}

\begin{center}
\fbox{\begin{minipage}[t]{0.97\columnwidth}%
\includegraphics[width=0.99\columnwidth]{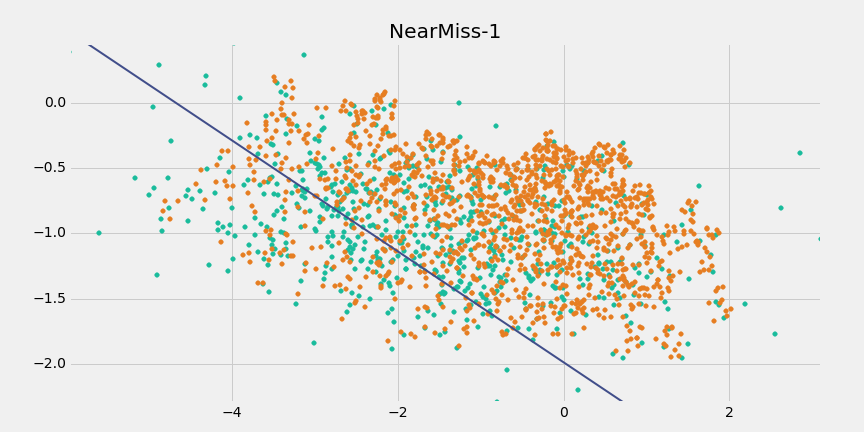}%
\end{minipage}}
\par\end{center}

\begin{center}
\begin{tabular}{|c|c|}
\hline 
precision on $L$ & recall on $S$\tabularnewline
\hline 
\hline 
0.92 & 0.32\tabularnewline
\hline 
\end{tabular}
\par\end{center}

\subsubsection{NearMiss-2}

In contrast to NearMiss-1, NearMiss-2 keeps those points from $L$
whose mean distance to the $k$ farthest points in $S$ is lowest.
\begin{center}
\fbox{\begin{minipage}[t]{0.97\columnwidth}%
\includegraphics[width=0.99\columnwidth]{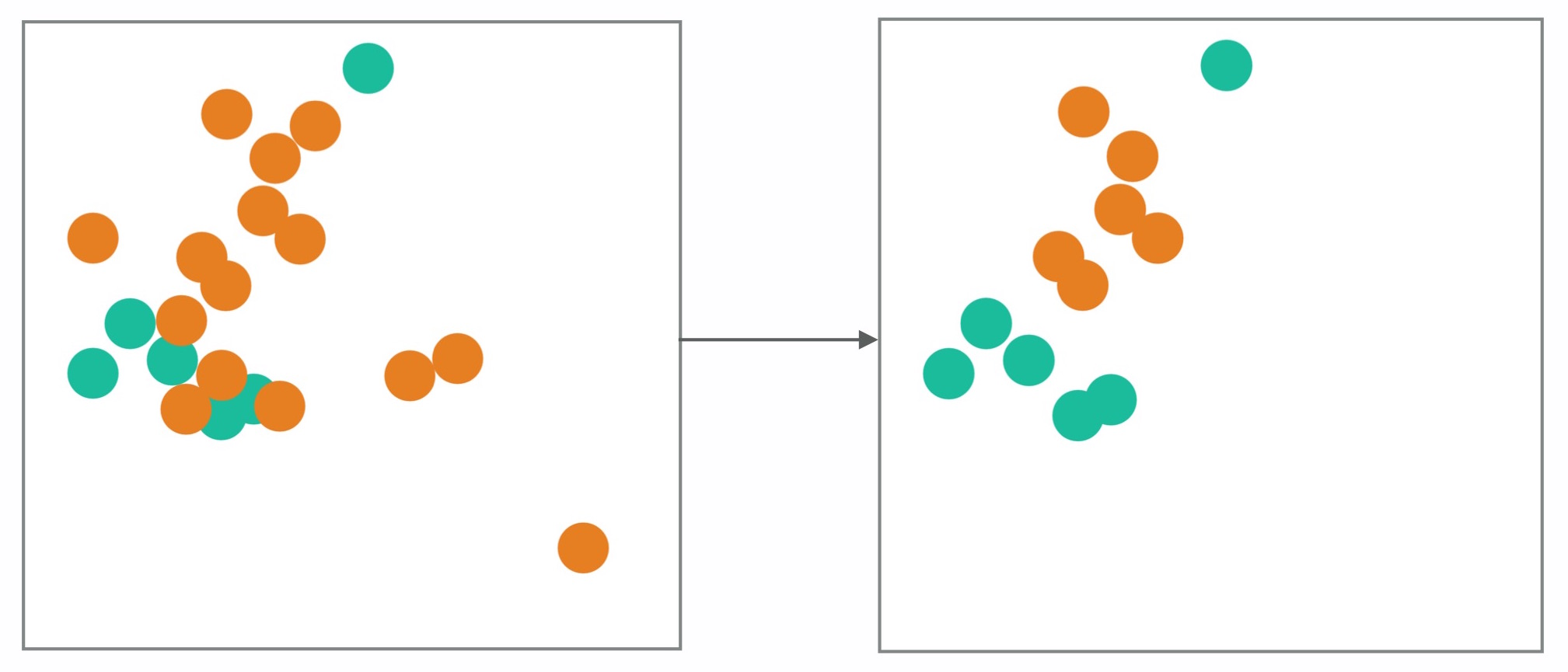}%
\end{minipage}}
\par\end{center}

We show below the result of using NearMiss-2 with $k=3$.
\begin{center}
\begin{tabular}{|c|c|c|}
\hline 
 & $|L|$ & $|S|$\tabularnewline
\hline 
\hline 
Before resampling & 6320 & 680\tabularnewline
\hline 
After resampling & 1360 & 680\tabularnewline
\hline 
\end{tabular}
\par\end{center}

\begin{center}
\fbox{\begin{minipage}[t]{0.97\columnwidth}%
\includegraphics[width=0.99\columnwidth]{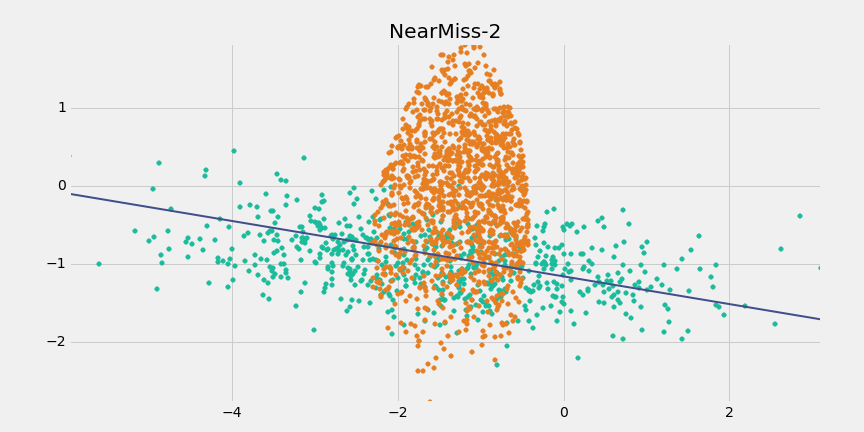}%
\end{minipage}}
\par\end{center}

\begin{center}
\begin{tabular}{|c|c|}
\hline 
precision on $L$ & recall on $S$\tabularnewline
\hline 
\hline 
0.95 & 0.60\tabularnewline
\hline 
\end{tabular}
\par\end{center}

\subsection{NearMiss-3}

The final NearMiss variant, NearMiss-3 selects $k$ nearest neighbors
in $L$ for every point in $S$. In this case, the undersampling ratio
is directly controlled by $k$ and is not separately tuned.
\begin{center}
\fbox{\begin{minipage}[t]{0.97\columnwidth}%
\includegraphics[width=0.99\columnwidth]{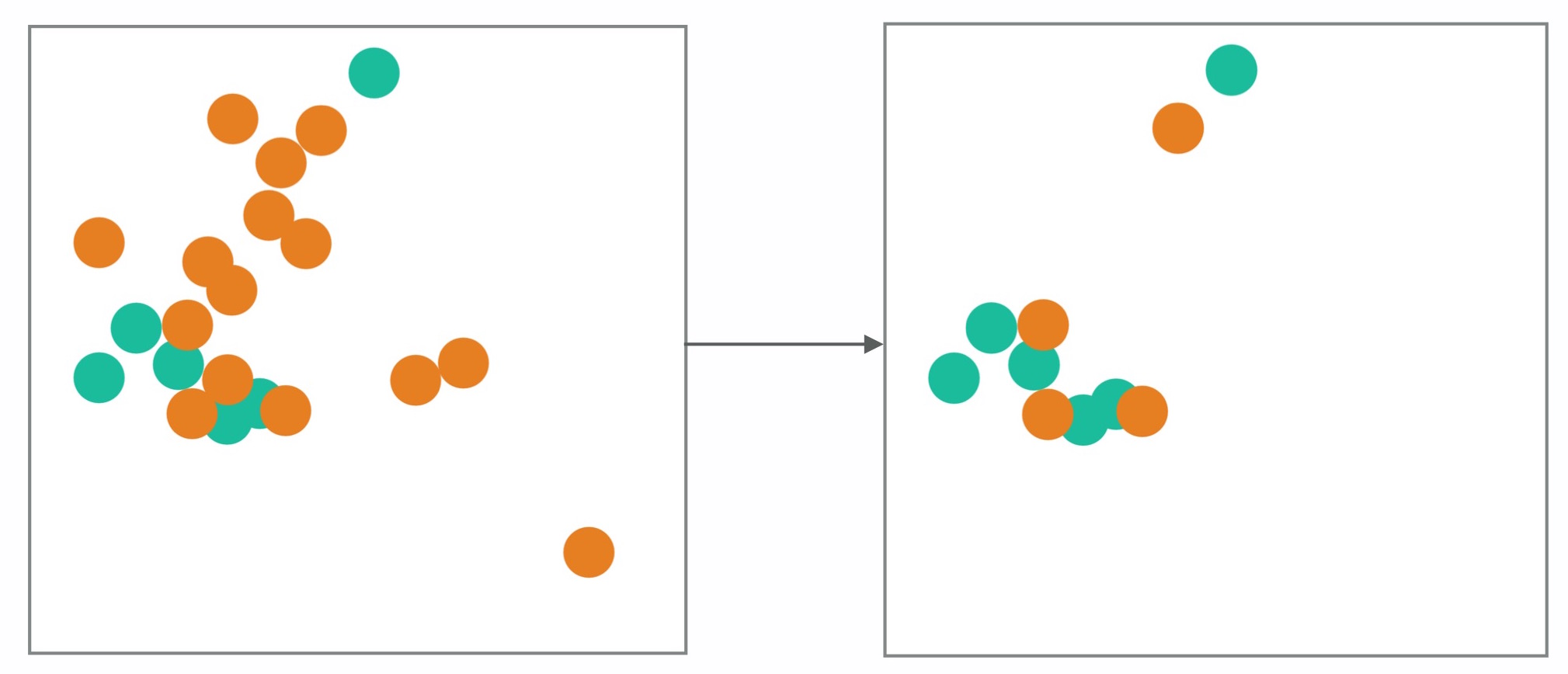}%
\end{minipage}}
\par\end{center}

We show below the result of using NearMiss-3 with $k=3$.
\begin{center}
\begin{tabular}{|c|c|c|}
\hline 
 & $|L|$ & $|S|$\tabularnewline
\hline 
\hline 
Before resampling & 6320 & 680\tabularnewline
\hline 
After resampling & 964 & 680\tabularnewline
\hline 
\end{tabular}
\par\end{center}

\begin{center}
\fbox{\begin{minipage}[t]{0.97\columnwidth}%
\includegraphics[width=0.99\columnwidth]{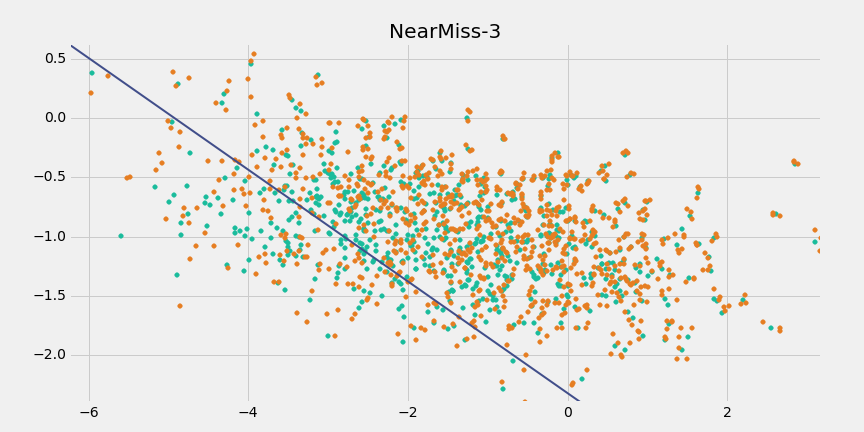}%
\end{minipage}}
\par\end{center}

\begin{center}
\begin{tabular}{|c|c|}
\hline 
precision on $L$ & recall on $S$\tabularnewline
\hline 
\hline 
0.91 & 0.20\tabularnewline
\hline 
\end{tabular}
\par\end{center}

\subsubsection{Condensed Nearest Neighbor (CNN)}

In CNN undersampling \cite{cnn}, the goal is to choose a subset $U$
of the training set $T$ such that for every point in $T$ its nearest
neighbor in $U$ is of the same class. $U$ can be grown iteratively
as follows:

\medskip{}

\noindent\fbox{\begin{minipage}[t]{1\columnwidth - 2\fboxsep - 2\fboxrule}%
\begin{enumerate}
\item Select a random point from $T$ and set $U=\{p\}$.
\item Scan $T-U$ and add to $U$ the first point found whose nearest neighbor
in $U$ is of a different class 
\item Repeat step 2 until $U$ is maximal
\end{enumerate}
\end{minipage}}

\medskip{}

Undersampling via CNN can be slower compared to other methods since
it requires many passes over the training data. Further, because of
the randomness involved in the selection of points at each iteration,
the subset selected can vary significantly.
\begin{center}
\fbox{\begin{minipage}[t]{0.97\columnwidth}%
\includegraphics[width=0.99\columnwidth]{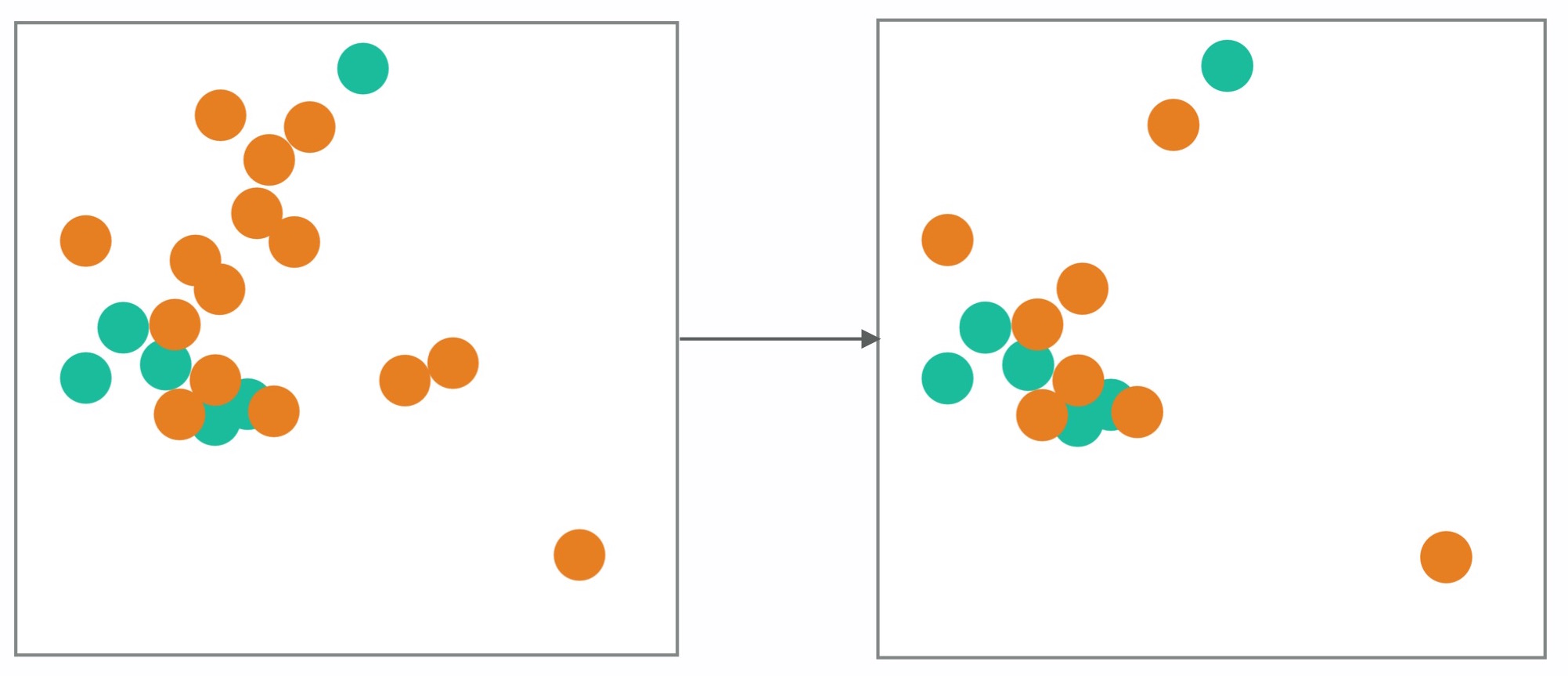}%
\end{minipage}}
\par\end{center}

A variant of CNN is to only undersample $L$ \emph{i.e. }retain all
points from $S$ but retain only those points in $L$ that belong
to $U$. We show performance below using this variant.
\begin{center}
\begin{tabular}{|c|c|c|}
\hline 
 & $|L|$ & $|S|$\tabularnewline
\hline 
\hline 
Before resampling & 6320 & 680\tabularnewline
\hline 
After resampling & 882 & 680\tabularnewline
\hline 
\end{tabular}
\par\end{center}

\begin{center}
\fbox{\begin{minipage}[t]{0.97\columnwidth}%
\includegraphics[width=0.99\columnwidth]{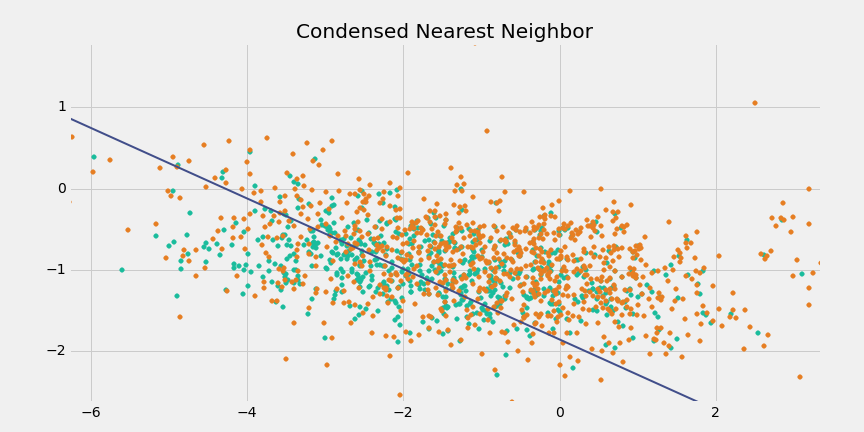}%
\end{minipage}}
\par\end{center}

\begin{center}
\begin{tabular}{|c|c|}
\hline 
precision on $L$ & recall on $S$\tabularnewline
\hline 
\hline 
0.93 & 0.39\tabularnewline
\hline 
\end{tabular}
\par\end{center}

\subsubsection{Edited Nearest Neighbor (ENN)}

In ENN \cite{enn}, undersampling of the majority class is done by
removing points whose class label differs from a majority of its $k$
nearest neighbors. 
\begin{center}
\fbox{\begin{minipage}[t]{0.97\columnwidth}%
\includegraphics[width=0.99\columnwidth]{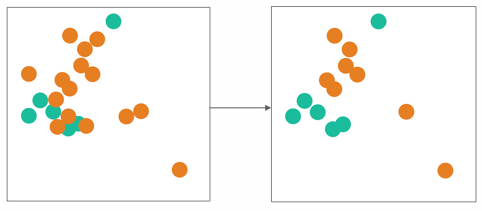}%
\end{minipage}}
\par\end{center}

The following results were obtained by employing ENN with $k=5$.
\begin{center}
\begin{tabular}{|c|c|c|}
\hline 
 & $|L|$ & $|S|$\tabularnewline
\hline 
\hline 
Before resampling & 6320 & 680\tabularnewline
\hline 
After resampling & 5120 & 680\tabularnewline
\hline 
\end{tabular}
\par\end{center}

\begin{center}
\fbox{\begin{minipage}[t]{0.97\columnwidth}%
\includegraphics[width=0.99\columnwidth]{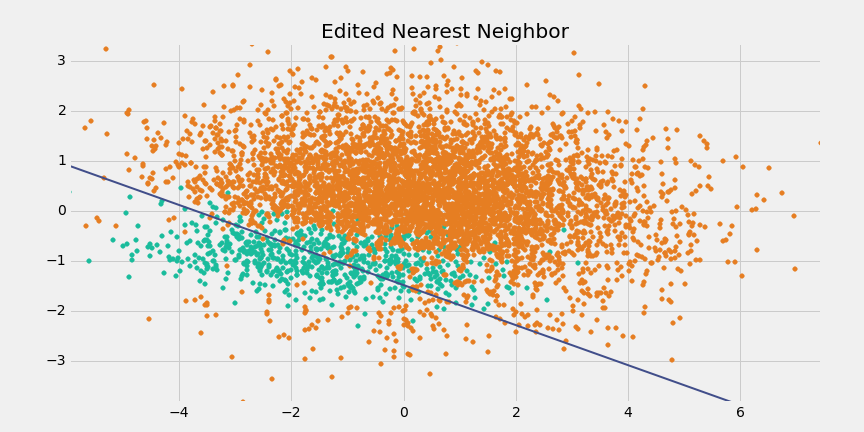}%
\end{minipage}}
\par\end{center}

\begin{center}
\begin{tabular}{|c|c|}
\hline 
precision on $L$ & recall on $S$\tabularnewline
\hline 
\hline 
0.95 & 0.59\tabularnewline
\hline 
\end{tabular}
\par\end{center}

\subsubsection{Repeated Edited Nearest Neighbor }

In Repeated Edited Nearest Neighbor, the ENN algorithm is applied
successively until ENN can remove no further points.
\begin{center}
\begin{tabular}{|c|c|c|}
\hline 
 & $|L|$ & $|S|$\tabularnewline
\hline 
\hline 
Before resampling & 6320 & 680\tabularnewline
\hline 
After resampling & 4796 & 680\tabularnewline
\hline 
\end{tabular}
\par\end{center}

\begin{center}
\fbox{\begin{minipage}[t]{0.97\columnwidth}%
\includegraphics[width=0.99\columnwidth]{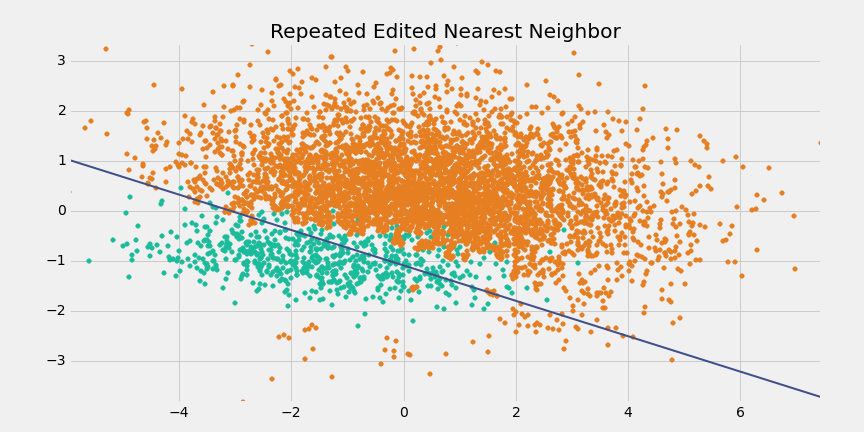}%
\end{minipage}}
\par\end{center}

\begin{center}
\begin{tabular}{|c|c|}
\hline 
precision on $L$ & recall on $S$\tabularnewline
\hline 
\hline 
0.97 & 0.80\tabularnewline
\hline 
\end{tabular}
\par\end{center}

\subsubsection{Tomek Link Removal}

A pair of examples is called a Tomek link if they belong to different
classes and are each other\textquoteright s nearest neighbors \cite{tomek1976two}.
Undersampling can be done by removing all tomek links from the dataset.
An alternate method is to only remove the majority class samples that
are part of a Tomek link. 
\begin{center}
\fbox{\begin{minipage}[t]{0.97\columnwidth}%
\includegraphics[width=0.99\columnwidth]{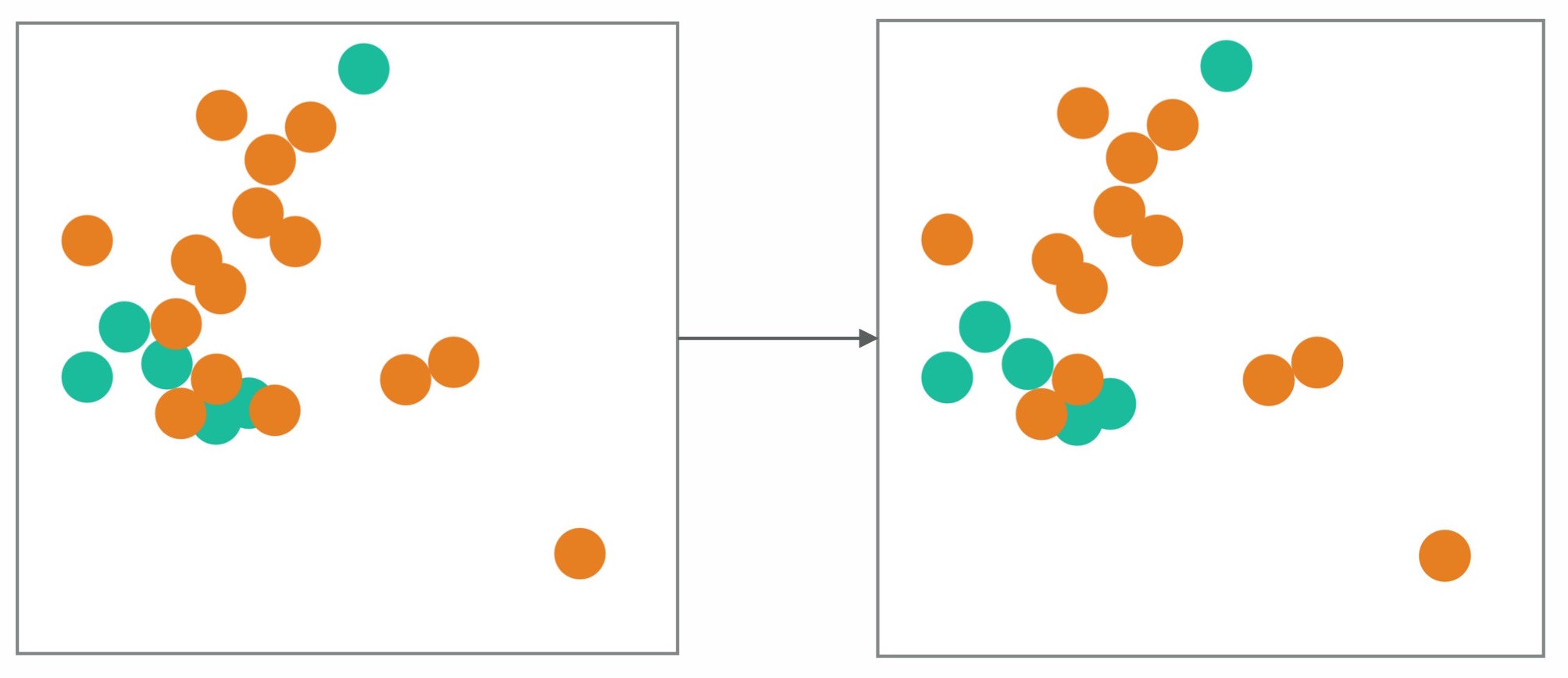}%
\end{minipage}}
\par\end{center}

We show the performance of the latter technique below.
\begin{center}
\begin{tabular}{|c|c|c|}
\hline 
 & $|L|$ & $|S|$\tabularnewline
\hline 
\hline 
Before resampling & 6320 & 680\tabularnewline
\hline 
After resampling & 6051 & 680\tabularnewline
\hline 
\end{tabular}
\par\end{center}

\begin{center}
\fbox{\begin{minipage}[t]{0.97\columnwidth}%
\includegraphics[width=0.99\columnwidth]{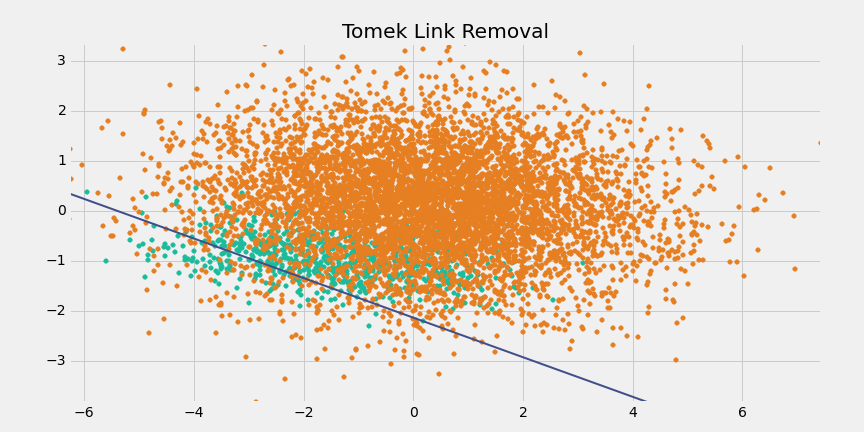}%
\end{minipage}}
\par\end{center}

\begin{center}
\begin{tabular}{|c|c|}
\hline 
precision on $L$ & recall on $S$\tabularnewline
\hline 
\hline 
0.91 & 0.21\tabularnewline
\hline 
\end{tabular}
\par\end{center}

\subsection{Oversampling methods}

At the other end of the spectrum are methods oversampling points from
the minority class. We explore a few such methods in this section.

\subsubsection{Random oversampling of minority class}

Points from the minority class may be oversampled with replacement.
This method is prone to overfitting. We consider the result of oversampling
of $S$ to achieve $r=0.5$.
\begin{center}
\begin{tabular}{|c|c|c|}
\hline 
 & $|L|$ & $|S|$\tabularnewline
\hline 
\hline 
Before resampling & 6320 & 680\tabularnewline
\hline 
After resampling & 6320 & 3160\tabularnewline
\hline 
\end{tabular}
\par\end{center}

\begin{center}
\fbox{\begin{minipage}[t]{0.97\columnwidth}%
\includegraphics[width=0.99\columnwidth]{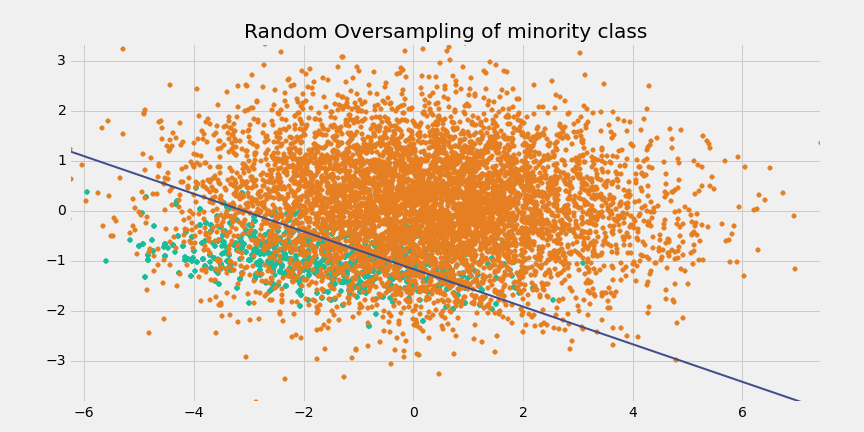}%
\end{minipage}}
\par\end{center}

\begin{center}
\begin{tabular}{|c|c|}
\hline 
precision on $L$ & recall on $S$\tabularnewline
\hline 
\hline 
0.97 & 0.76\tabularnewline
\hline 
\end{tabular}
\par\end{center}

\subsubsection{SMOTE}

A more sophisticated means for oversampling is Synthetic Minority
Oversampling Technique (SMOTE) which is outlined below \cite{chawla2002smote}.

\medskip{}

\noindent\fbox{\begin{minipage}[t]{1\columnwidth - 2\fboxsep - 2\fboxrule}%
For each point $p$ in $S$:
\begin{enumerate}
\item Compute its $k$ nearest neighbors in $S$.
\item Randomly choose $r\le k$ of the neighbors (with replacement).
\item Choose a random point along the lines joining $p$ and each of the
$r$ selected neighbors.
\item Add these synthetic points to the dataset with class $S$.
\end{enumerate}
\end{minipage}}

\medskip{}

We show below the results of applying SMOTE with $k=5$ in order to
achieve $r=0.5$.
\begin{center}
\fbox{\begin{minipage}[t]{0.97\columnwidth}%
\includegraphics[width=0.99\columnwidth]{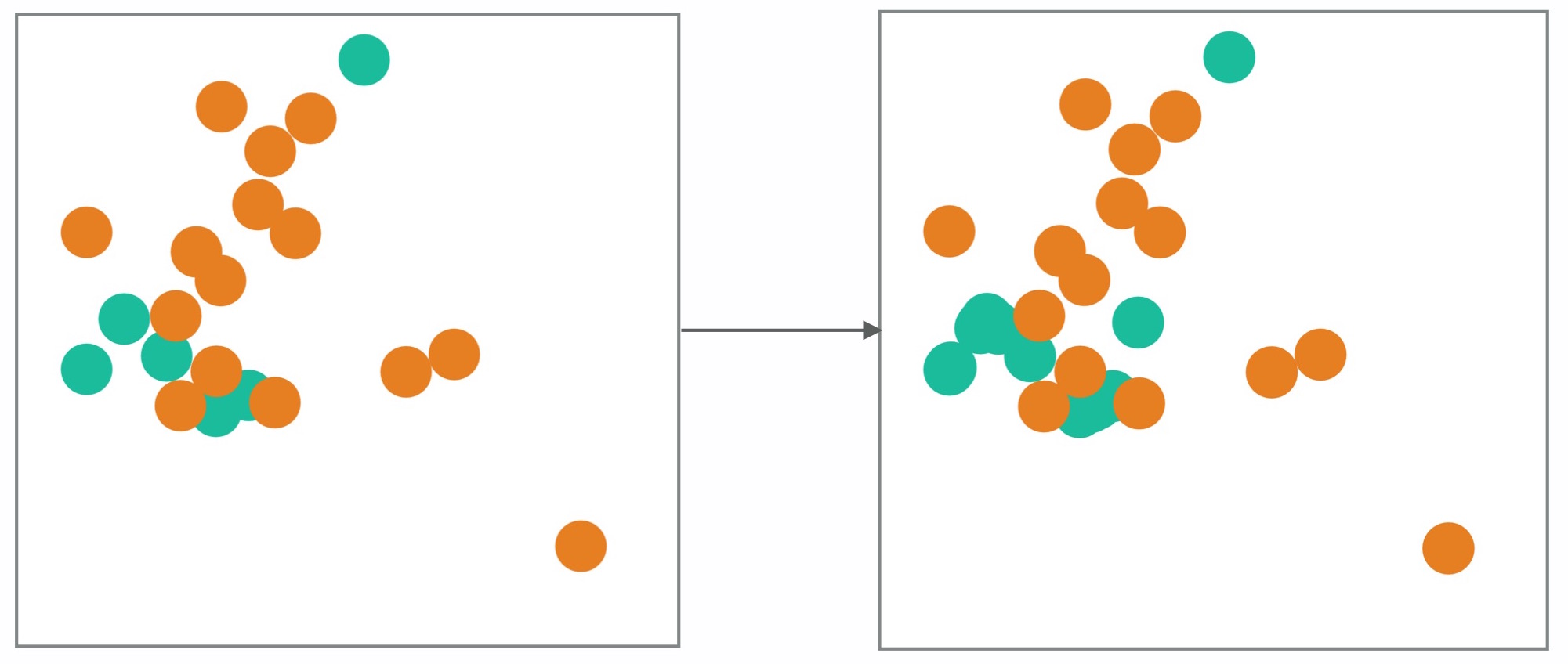}%
\end{minipage}}
\par\end{center}

\begin{center}
\begin{tabular}{|c|c|c|}
\hline 
 & $|L|$ & $|S|$\tabularnewline
\hline 
\hline 
Before resampling & 6320 & 680\tabularnewline
\hline 
After resampling & 6320 & 3160\tabularnewline
\hline 
\end{tabular}
\par\end{center}

\begin{center}
\fbox{\begin{minipage}[t]{0.97\columnwidth}%
\includegraphics[width=0.99\columnwidth]{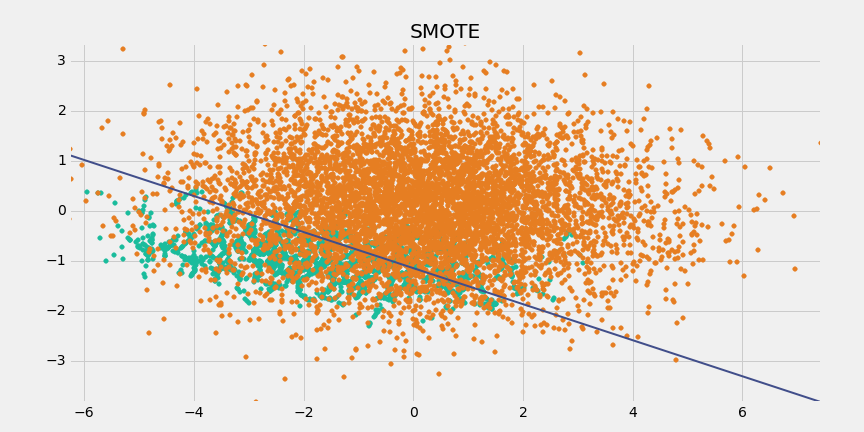}%
\end{minipage}}
\par\end{center}

\begin{center}
\begin{tabular}{|c|c|}
\hline 
precision on $L$ & recall on $S$\tabularnewline
\hline 
\hline 
0.97 & 0.77\tabularnewline
\hline 
\end{tabular}
\par\end{center}

\subsubsection{Borderline-SMOTE1}

There are two enhancements of SMOTE, termed borderline SMOTE \cite{han2005borderline},
which may yield better performance than vanilla SMOTE. 

\medskip{}

\noindent\fbox{\begin{minipage}[t]{1\columnwidth - 2\fboxsep - 2\fboxrule}%
For each point $p$ in $S$:
\begin{enumerate}
\item Compute its $m$ nearest neighbors in $T$. Call this set $M_{p}$
and let $m^{'}=|M_{p}\cap L|$.
\item If $m^{'}=m$, $p$ is a noisy example. Ignore $p$ and continue to
the next point.
\item If $0\leq m^{'}\leq\frac{m}{2}$, $p$ is safe. Ignore $p$ and continue
to the next point.
\item If $\frac{m}{2}\leq m^{'}\leq m$, add $p$ to the set DANGER.
\end{enumerate}
For each point $d$ in DANGER, apply the SMOTE algorithm to generate
synthetic examples.%
\end{minipage}}

\medskip{}

We apply Borderline-SMOTE1 with $k=5$ in order to achieve $r=0.5$.
\begin{center}
\begin{tabular}{|c|c|c|}
\hline 
 & $|L|$ & $|S|$\tabularnewline
\hline 
\hline 
Before resampling & 6320 & 680\tabularnewline
\hline 
After resampling & 6320 & 3160\tabularnewline
\hline 
\end{tabular}
\par\end{center}

\begin{center}
\fbox{\begin{minipage}[t]{0.97\columnwidth}%
\includegraphics[width=0.99\columnwidth]{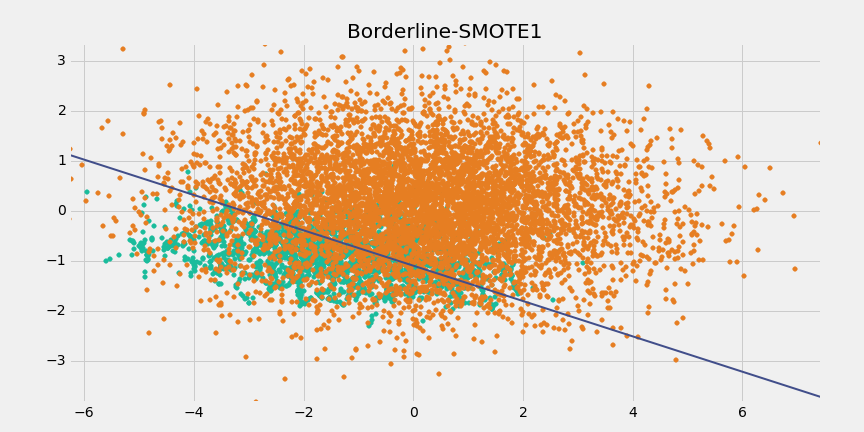}%
\end{minipage}}
\par\end{center}

\begin{center}
\begin{tabular}{|c|c|}
\hline 
precision on $L$ & recall on $S$\tabularnewline
\hline 
\hline 
0.97 & 0.78\tabularnewline
\hline 
\end{tabular}
\par\end{center}

\subsubsection{Borderline-SMOTE2}

Borderline-SMOTE2 is similar to Borderline-SMOTE1 except in the last
step, new synthetic examples are created along the line joining points
in DANGER to either their nearest neighbors in $S$ or their nearest
neighbors in $L$. In the latter case, the synthetic points are chosen
such that they are closer to the original point.

We apply Borderline-SMOTE2 with $k=5$ in order to achieve $r=0.5$.
\begin{center}
\begin{tabular}{|c|c|c|}
\hline 
 & $|L|$ & $|S|$\tabularnewline
\hline 
\hline 
Before resampling & 6320 & 680\tabularnewline
\hline 
After resampling & 6320 & 3160\tabularnewline
\hline 
\end{tabular}
\par\end{center}

\begin{center}
\fbox{\begin{minipage}[t]{0.97\columnwidth}%
\includegraphics[width=0.99\columnwidth]{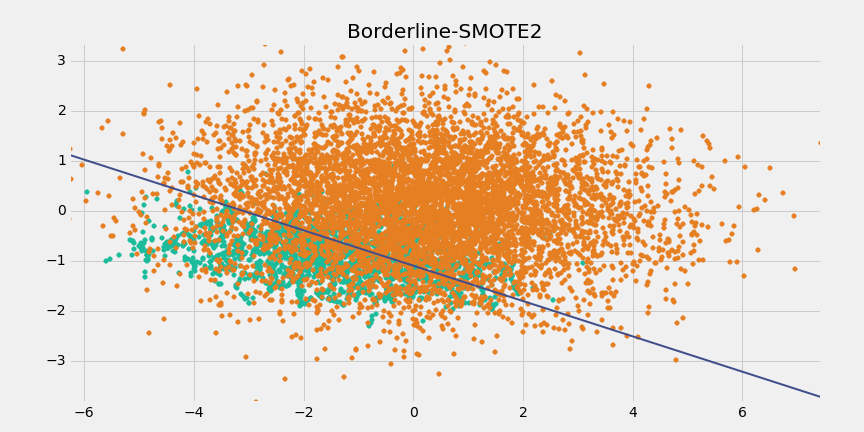}%
\end{minipage}}
\par\end{center}

\begin{center}
\begin{tabular}{|c|c|}
\hline 
precision on $L$ & recall on $S$\tabularnewline
\hline 
\hline 
0.97 & 0.80\tabularnewline
\hline 
\end{tabular}
\par\end{center}

\subsection{Combination methods}

Performing a combination of oversampling and undersampling can often
yield better results than either in isolation. We discuss two particular
combinations here.

\subsubsection{SMOTE + Tomek Link Removal}

We show the result of performing SMOTE with $k=5$ and $r=0.5$ followed
by Tomek link removal.
\begin{center}
\begin{tabular}{|c|c|c|}
\hline 
 & $|L|$ & $|S|$\tabularnewline
\hline 
\hline 
Before resampling & 6320 & 680\tabularnewline
\hline 
After resampling & 6050 & 3160\tabularnewline
\hline 
\end{tabular}
\par\end{center}

\begin{center}
\fbox{\begin{minipage}[t]{0.97\columnwidth}%
\includegraphics[width=0.99\columnwidth]{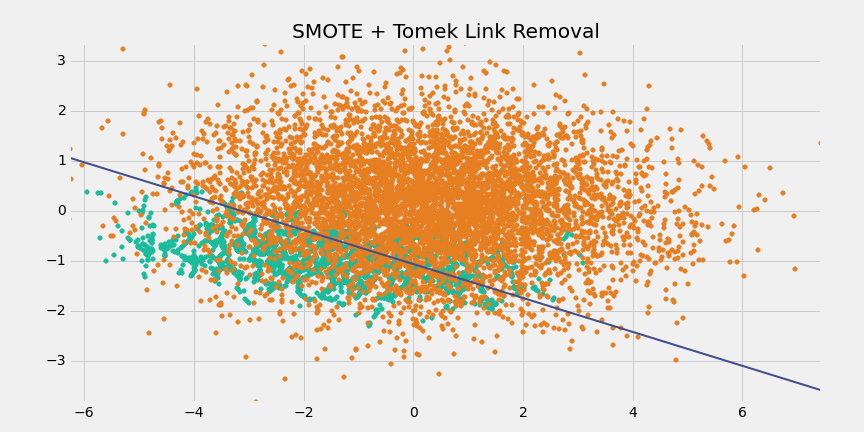}%
\end{minipage}}
\par\end{center}

\begin{center}
\begin{tabular}{|c|c|}
\hline 
precision on $L$ & recall on $S$\tabularnewline
\hline 
\hline 
0.97 & 0.80\tabularnewline
\hline 
\end{tabular}
\par\end{center}

\subsubsection{SMOTE + ENN}

The following result is obtained by performing SMOTE with $k=5$ and
$r=0.5$ followed by ENN with $k=5$.
\begin{center}
\begin{tabular}{|c|c|c|}
\hline 
 & $|L|$ & $|S|$\tabularnewline
\hline 
\hline 
Before resampling & 6320 & 680\tabularnewline
\hline 
After resampling & 4894 & 3160\tabularnewline
\hline 
\end{tabular}
\par\end{center}

\begin{center}
\fbox{\begin{minipage}[t]{0.97\columnwidth}%
\includegraphics[width=0.99\columnwidth]{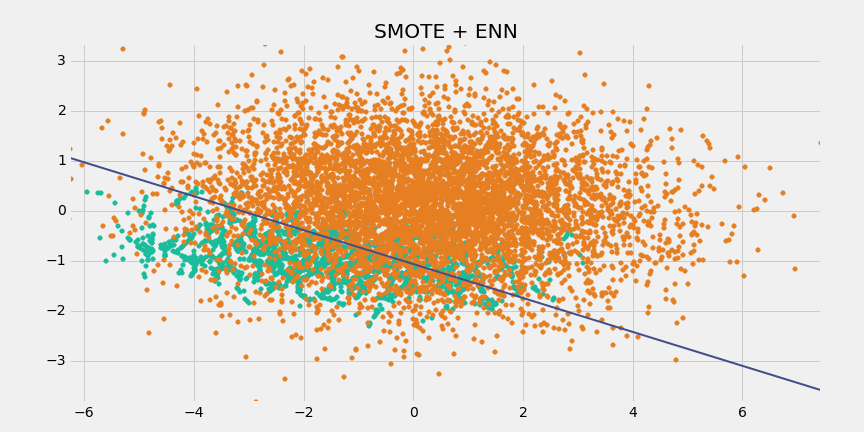}%
\end{minipage}}
\par\end{center}

\begin{center}
\begin{tabular}{|c|c|}
\hline 
precision on $L$ & recall on $S$\tabularnewline
\hline 
\hline 
0.97 & 0.92\tabularnewline
\hline 
\end{tabular}
\par\end{center}

\subsection{Ensemble methods}

\subsubsection{EasyEnsemble}

In EasyEnsemble \cite{ensemble} a sequence of classifiers are built
by resampling the majority class. The algorithm is outlined below.

\noindent\fbox{\begin{minipage}[t]{1\columnwidth - 2\fboxsep - 2\fboxrule}%
\begin{enumerate}
\item For $i=1,...,N$:
\begin{enumerate}
\item Randomly sample a subset $L_{i}$ of $L$ such that $|L_{i}|=|S|$.
\item Learn an AdaBoost ensemble using $L_{i}$ and $S$
\[
F_{i}(x)=sgn(\Sigma_{j=1}^{n_{i}}w_{ij}f_{ij}(x)-b_{i})
\]
\end{enumerate}
\item Combine the above classifiers into a meta-ensemble
\[
F(x)=sgn(\Sigma_{i=1}^{N}(\Sigma_{j=1}^{n_{i}}w_{ij}f_{ij}(x)-b_{i}))
\]
\end{enumerate}
\end{minipage}}
\begin{center}
\begin{tabular}{|c|c|}
\hline 
precision on $L$ & recall on $S$\tabularnewline
\hline 
\hline 
0.98 & 0.88\tabularnewline
\hline 
\end{tabular}
\par\end{center}

\subsubsection{BalanceCascade}

BalanceCascade \cite{ensemble} is similar to EasyEnsemble except
the classifier created in each iteration influences the selection
of points in the next iteration.

\noindent\fbox{\begin{minipage}[t]{1\columnwidth - 2\fboxsep - 2\fboxrule}%
\begin{enumerate}
\item Set $t=r^{\frac{1}{N-1}}$
\item For $i=1,...,N$:
\begin{enumerate}
\item Randomly sample a subset $L_{i}$ of $L$ such that $|L_{i}|=|S|$.
\item Learn an AdaBoost ensemble using $L_{i}$ and $S$
\[
F_{i}(x)=sgn(\Sigma_{j=1}^{n_{i}}w_{ij}f_{ij}(x)-b_{i})
\]
\item Tune $b_{i}$ such that the false positive rate for $F_{i}$ is $t$.
\end{enumerate}
\item Undersample $L$ to remove points correctly classified by $F_{i}$.
\item Combine the above classifiers into a meta-ensemble
\[
F(x)=sgn(\Sigma_{i=1}^{N}(\Sigma_{j=1}^{n_{i}}w_{ij}f_{ij}(x)-b_{i}))
\]
\end{enumerate}
\end{minipage}}
\begin{center}
\begin{tabular}{|c|c|}
\hline 
precision on $L$ & recall on $S$\tabularnewline
\hline 
\hline 
0.99 & 0.91\tabularnewline
\hline 
\end{tabular}
\par\end{center}

\section{Conclusion}

In this paper we discussed some resampling techniques to improve classification
performance on the minority class in the presence of data imbalance.
We presented the performance of several methods on a synthetic dataset
in terms of precision on the majority class and recall on the minority
class. 

The methods discussed in this paper are by no means an exhaustive
list. Several other techqniues have been proposed in literature which
have had success in handling data imbalance. Some of these include
One side selection \cite{kubat1997addressing}, ADASYN \cite{he2008adasyn},
SVM SMOTE \cite{nguyen2011borderline}, SMOTEBoost \cite{chawla2003smoteboost},
Cluster-Based Oversampling \cite{cbo}, Kernel-based methods and active
learning \cite{he2009learning}.

On our synthetic dataset, with respect to our chosen metric, the methods
SMOTE+ENN in combination with a logistic regression classifier and
BalanceCascade give the best performance. However, depending on the
data distribution, the presence of within class imbalance in addition
to between class imbalance and the choice of classifier used on resampled
datasets, other methods may yield better results.

\bibliographystyle{plain}
\bibliography{references}

\begin{thebibliography}{10}

\bibitem{chawla2002smote}
Nitesh~V. Chawla, Kevin~W. Bowyer, Lawrence~O. Hall, and W.~Philip Kegelmeyer.
\newblock Smote: synthetic minority over-sampling technique.
\newblock {\em Journal of artificial intelligence research}, 16:321--357, 2002.

\bibitem{chawla2003smoteboost}
Nitesh~V Chawla, Aleksandar Lazarevic, Lawrence~O Hall, and Kevin~W Bowyer.
\newblock Smoteboost: Improving prediction of the minority class in boosting.
\newblock In {\em European Conference on Principles of Data Mining and
  Knowledge Discovery}, pages 107--119. Springer, 2003.

\bibitem{han2005borderline}
Hui Han, Wen-Yuan Wang, and Bing-Huan Mao.
\newblock Borderline-smote: a new over-sampling method in imbalanced data sets
  learning.
\newblock In {\em International Conference on Intelligent Computing}, pages
  878--887. Springer, 2005.

\bibitem{cnn}
P.~Hart.
\newblock The condensed nearest neighbor rule (corresp.).
\newblock {\em IEEE Trans. Inf. Theor.}, 14(3):515--516, September 2006.

\bibitem{he2008adasyn}
Haibo He, Yang Bai, Edwardo~A Garcia, and Shutao Li.
\newblock Adasyn: Adaptive synthetic sampling approach for imbalanced learning.
\newblock In {\em 2008 IEEE International Joint Conference on Neural Networks
  (IEEE World Congress on Computational Intelligence)}, pages 1322--1328. IEEE,
  2008.

\bibitem{he2009learning}
Haibo He and Edwardo~A Garcia.
\newblock Learning from imbalanced data.
\newblock {\em IEEE Transactions on knowledge and data engineering},
  21(9):1263--1284, 2009.

\bibitem{cbo}
Taeho Jo and Nathalie Japkowicz.
\newblock Class imbalances versus small disjuncts.
\newblock {\em ACM Sigkdd Explorations Newsletter}, 6(1):40--49, 2004.

\bibitem{king2001logistic}
Gary King and Langche Zeng.
\newblock Logistic regression in rare events data.
\newblock {\em Political analysis}, 9(2):137--163, 2001.

\bibitem{kubat1997addressing}
Miroslav Kubat, Stan Matwin, et~al.
\newblock Addressing the curse of imbalanced training sets: one-sided
  selection.
\newblock In {\em ICML}, volume~97, pages 179--186. Nashville, USA, 1997.

\bibitem{ensemble}
Xu-Ying Liu, Jianxin Wu, and Zhi-Hua Zhou.
\newblock Exploratory undersampling for class-imbalance learning.
\newblock {\em IEEE Transactions on Systems, Man, and Cybernetics, Part B
  (Cybernetics)}, 39(2):539--550, 2009.

\bibitem{nm}
Inderjeet Mani and I~Zhang.
\newblock knn approach to unbalanced data distributions: a case study involving
  information extraction.
\newblock In {\em Proceedings of workshop on learning from imbalanced
  datasets}, 2003.

\bibitem{nguyen2011borderline}
Hien~M Nguyen, Eric~W Cooper, and Katsuari Kamei.
\newblock Borderline over-sampling for imbalanced data classification.
\newblock {\em International Journal of Knowledge Engineering and Soft Data
  Paradigms}, 3(1):4--21, 2011.

\bibitem{tomek1976two}
Ivan Tomek.
\newblock Two modifications of cnn.
\newblock {\em IEEE Trans. Systems, Man and Cybernetics}, 6:769--772, 1976.

\bibitem{enn}
Dennis~L Wilson.
\newblock Asymptotic properties of nearest neighbor rules using edited data.
\newblock {\em IEEE Transactions on Systems, Man, and Cybernetics},
  (3):408--421, 1972.

\end{thebibliography}

\end{document}